# When elephants nodded and dolls spoke: Bringing together robotics and storytelling for environmental literacy

Mukil M.V.
AMMACHI Labs
Amrita Vishwa Vidyapeetham
Amritapuri, India
mukil.mv@ammachilabs.org

Gayathri Manikutty
AMMACHI Labs
Amrita Vishwa Vidyapeetham
Amritapuri, India
gayathrim@am.amrita.edu

Divya Vijayan
AMMACHI Labs
Amrita Vishwa Vidyapeetham
Amritapuri, India
divya.vijayan@ammachilabs.org

Aparna Rangudu
AMMACHI Labs
Amrita Vishwa Vidyapeetham
Amritapuri, India
aparna.rangudu@ammachilabs.org

Rao R. Bhavani
AMMACHI Labs
Amrita Vishwa Vidyapeetham
Amritapuri, India
bhavani@am.amrita.edu

## ABSTRACT

Inculcating principles of environmental stewardship among the children and youth is needed urgently today for creating a sustainable future. This paper presents a model for promoting environment literacy in India using story telling based workshops while focusing on STEM education including computational thinking, robotics and maker skills. During the workshop, participants build a robotic diorama with digital animations and animatronics to tell their story. Our initial observations from pilot studies conducted in 2019 in six rural and semi-urban schools in India showed us that the children were deeply engaged and enthusiastic throughout the workshop making the entire learning experience a very meaningful and joyful one for all.

## CCS CONCEPTS

• Applied computing → Interactive learning environments; • Applied computing → Collaborative learning; • Social and professional topics → K-12 education; • Social and professional topics → Computational thinking

## KEYWORDS

Experiential learning; 21st century skills; middle school education; co-creation; K-12 education; Robotics

## 1 Introduction

The world today is facing unprecedented challenges such as climate change, food security, rapid technological advances, conflicts, and global health epidemics, which need to be dealt with both at global and local levels [1]. One cannot predict exactly how these challenges will compound over time and what severity levels will they attain in the next decade but, certainly, human's environmental actions will have a central role in determining how safe our world would be for humans to live. Therefore children must be taught from an early age to think critically and reflect deeply about how each of our decisions and actions impact the environment and our daily lives [2].

Whereas environmental science has become an integral part of the K-12 curriculum in schools in India, there is a lot of emphasis on presenting factual data and little emphasis on bringing inter-disciplinary experiential learning into the classroom. Hence the learning becomes that of rote-learning with the child often missing the critical link between the subject and his/her daily life [3]. The primary objective of our research work was to introduce middle school children to experiential science and technology based learning for environment literacy using robotics and maker skills that is situated within the lives of the children. The secondary objective was to promote 21$^{st}$ century skills of creativity, collaboration, communication and critical thinking for environment literacy. By the term environmental literacy we are referring to children developing knowledge and understanding of different environment problems, developing skills and motivations for spreading awareness on the issues that they empathize with and developing innovative solutions for the environmental problems they are championing [4, 5].

To be successful in this technological age, education must be such that it encourages youth to be creators of technology and not just consumers of technology [6]. Experiential learning of Science, Technology, Engineering, Arts and Mathematics (STEAM) subjects through robotics and maker skills could be one way to impart technology education. Our previous work with middle school children [7] and that done by our colleagues Sushmita et. al [8] highlighted that experiential learning through maker skills and robotics provides a very engaging and positive learning experience that could potentially improve their self-efficacy and motivation. These studies were done with children who have little prior technology, robotics or maker experience.

We adopted a digital story telling based approach in our workshop as a means to integrate technology and environment literacy because children have a natural way of telling stories that are compelling and emotionally engaging. Also, the storytelling process itself encourages children to be creative and imaginative. Prior research with digital storytelling and making has shown that children have always discovered innovative ways to create meaningful interactions with their audience through story telling [9, 10].

The rest of the paper is organized as follows. Section II discusses methodology we adopted for our workshop and our initial observations. In Section III, we present some feedback from students, and discuss conclusions and future work.

## 2. Methodology and Observations

### 2.1 Instructional Setting

We have piloted the creative workshop for about 188 students in five schools in South India since January 2019. During the pilot phase, the curriculum was designed and refined with five different workshop classes in the five schools (approximately 40 students in each class). The classes were typically conducted for $7^{th}$ grade, $8^{th}$ grade and $9^{th}$ grade students who self-selected themselves to be part of the class. The students participated in the workshop class for 3 continuous days for 6 hours each day.

### 2.2 Workshop structure

The children began the creative robotics workshop class learning the basics of physical computing and robotics. They learnt about finite state machines (FSMs) and how they can construct these through tangible materials like paper by creating flexagons. They used the flexagons to create stories organized as scenes. From this activity, they moved on to making and learning electronics, including LEDs, sensors, and actuators - the nuts and bolts of robotics. They cut and assembled paper robots which they programmed using the Scratch programming language. This set the stage for them to move on to creative stage of the workshop, namely, creating the robotic diorama (a miniature exhibit with three-dimensional figures) to tell their story. We will now describe each of these sessions in detail.

### 2.3 Making Flexagons

Students made a tri hexa flexagon and hexa hexa flexagon by folding papers. Flexagons were developed by mathematicians Arthur H. Stone and the members of the flexagon committee in the year 1939 [11]. They have the property of changing their faces when they are 'flexed'. Students learnt about Tuckerman's diagram, a graphical representation of the finite state machine which explains the path to discover different faces of the tri or hexa hexa flexagons. After learning about flexagons and playing with it, the children were instructed to come up with creative ideas for using the flexagon for promoting awareness on environment protection and conservation. The children were given the option to either write stories or write positive messages on the tri-hexa flexagon such that each face of the flexagon represented one scene of their story or one message they want to convey and each flex would take the story to the next scene or the next message. Below is a story written by one of the workshop participant:

> **Scene1:** *Duke and his BFF (best friend) Vroom were hanging out after school. They lived in Faitown, a small town in vehicle island. Once it was time for Vroom to go, he bid goodbye and was on his way home. On his way he fell into a deep pit in which he found a long tunnel.*
> **Scene2:** *As Duke found his way out of the tunnel he reached Goodtown, where the state of the town was disgusting. Withered trees, dried up lakes, loads of trash and the list goes on and on. He, on this shocking sight bumps onto his cousin Buzz, who was a Goodtown citizen. He, like all other vehicles, emitted a lot of smoke and he was quite different from Duke as there was quite an age difference between them.*
> **Scene 3:** *On seeing Buzz emit large amount of $CO_2$ emissions he wanted to end it, but what could he do about it? Then an idea struck him. He advised his cousin, Buzz to inform all citizens to change engines at the cable garage and plant new saplings. The very next month Goodtown has finally lived up to its name where the town was as shiny and clean as it could ever have become.*

We will also share a message one of the participants depicted on their flexagon by drawing pictures and writing captions for the pictures:

> **Scene 1:** *A famous quote "No pollution is the only solution". It gives a message to all the people of the world that if we continue to pollute our environment like now then the end of the earth will come faster. So we now have only one solution - that is no pollution.*
> **Scene2:** *It shows the various things we can do to conserve the environment. Save the trees, do not cut the trees, say no to plastics they are very harmful. Follow the 3R's - Reduce, Reuse, Recycle, save the earth, save water for the future. Do not waste water.*
> **Scene 3:** *The third scene shows a very important and meaningful quote, "Environment is our need - don't destroy it for your greed." We need the environment but environment does not need us. So it our responsibility to conserve the environment for ourselves and for the future generations.*

We observed that the students grasped the concept of tri hexa flexagons transitions very easily since it had only three faces. The transitions of the hexa hexa flexagons were more challenging for the participants to understand and the Tuckerman diagram proved to be essential to explain the concept to them.

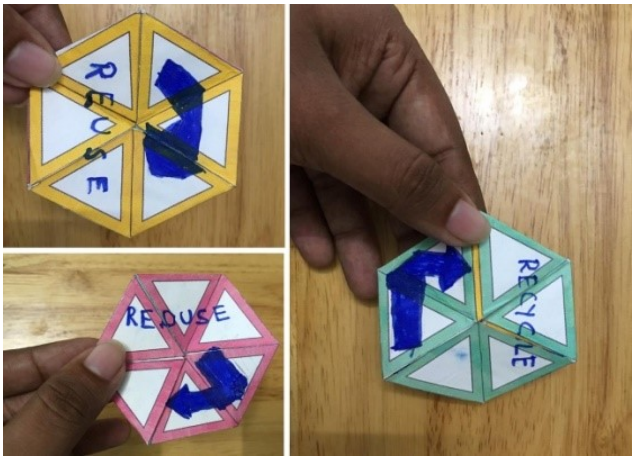

**Figure 1: Flexagon stories and messages on environment conservation**

## 2.4 Programing electrical and electronic components using Scratch

After making the flexagons, the children moved to the next phase where they used the Hummingbird robotics kit [12] and craft materials to learn the nuts and bolts of robotics. The Hummingbird kit is an educational robotics package that contains a controller, DC, servo and vibration motors, several sensors, and LED lights. The Hummingbird controller has clearly marked ports with component names and number so that the children can use an easy grip screwdriver for connecting components to the ports of the controller.

The children cut out the paper stencils and assembled them to create their paper puppets. They fixed motors and sensors to the paper puppets to make it move using the Scratch. Scratch [13] is a block based programming environment developed by Lifelong Kindergarten group at Massachusetts Institute of Technology that allows children aged 8-16 to create animations, interactive stories, games and music without needing to have prior knowledge of computation thinking. The blocks are shaped like puzzle pieces that interlock to create a program flow. Different block shapes introduce different concepts of computational thinking to the children.

The children had a prior workshop which introduced them to the basics of Scratch blocks. They built upon this knowledge and wrote Scratch programs with blocks to control the electronics components using the Hummingbird's Birdbrain interface. Fig. 2 shows an angry elephant robot puppet that half the class made and programmed using Scratch to shake its head up and down. The other half of the class programmed a paper doll to swivel.

After learning to program their puppets to move using sensors and actuators, the children moved on to learn about importing GIFs into Scratch and create short animation movie sequences using the imported images. They recorded their voices and imported music to overlay media on the animated sequences to create their own 5-10 second animated movies. By embedding the movement of puppets into their animated movies, the children were able to make their animated movies interactive.

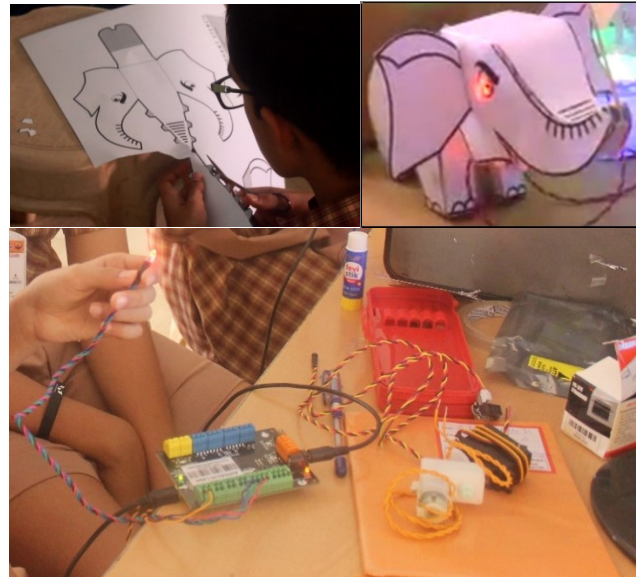

**Figure 2: Hummingbird controller connected with motors and sensors**.

## 2.5 Making the robotic diorama

For this session, the children were divided into three groups. Each group was asked to create a scene of a story with a robotic diorama. To seed their effort, they were shown a clip of how a robotic diorama could look like. They were given several cutouts of silhouettes to paint and create their scene with. Each group was asked to divide up the tasks of scene creation among themselves. The tasks involved painting the silhouettes and assembling it to setup the diorama, developing a story board and story narrative for their scene, creating several short animation sequences using GIFs to tell their story, recording voices and music for overlaying media on the animation sequences, and finally programming the paper puppets to add interactivity to the diorama. The children were asked the divide up the tasks among themselves through collaborative discussions within the group members. The children formed sub-groups for each of the tasks to work on it parallel. The constraints given to them were time (they had to put the diorama together in a day) and resources (they had to use the paper silhouettes and stationery given to them to create the diorama). They were strongly encouraged to be creative within the constraints given.

In each of the classes, we observed that the children divided up the tasks based on their personal preferences. Some sub-groups would expand upon their previously created animation sequences whereas others created new animation sequences. Almost all the groups painted and colored an elaborate diorama using card-stock paper and fixed it on a thermocol base for the story scenes. They added the computer screen as a backdrop for their diorama. Fig. 3 shows two scenes from a diorama that the participants built. The animation running on the backdrop added richness to the scene portrayed by the diorama. The paper puppets added interactivity to the diorama. The children created effects such as the paper puppet elephant nodding its head and trumpeting during an animated story narration and a paper doll swiveling to address the audience with questions on environment responsibilities for the audience to ponder about. The knowledge of FSMs allowed the participants to create story variations based on their interactions with the diorama.

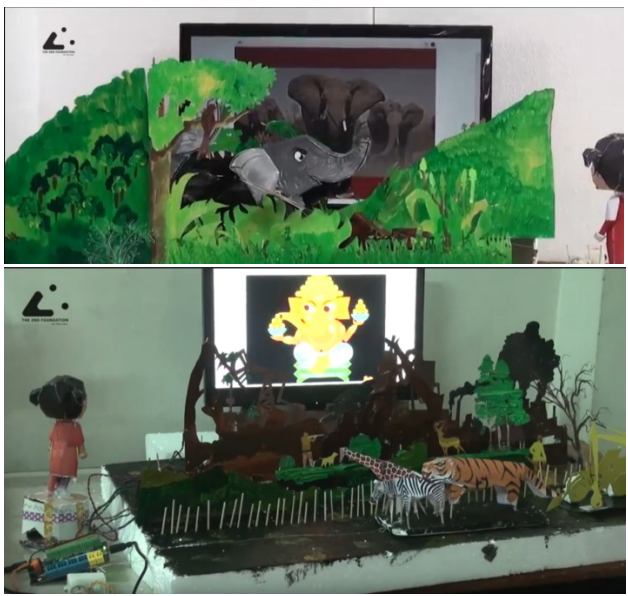

**Figure 3: Creative robotics diorama developed by students**.

This session brought together multiple 21$^{st}$ century skills from the student groups – artistic creativity in terms of painting and arranging the scenes of the diorama and programming LED light and music effects to enliven the scenes, critical thinking skills in terms of programming the background animation and the paper robots using the Hummingbird kit, collaboration in terms of dividing up the tasks involved in creating the diorama and learning to leverage each other's strengths to put it together and communication skills in terms of creating the message to convey through their diorama. We observed that the children created personalized stories based on their knowledge and understanding of environmental awareness. Most of the stories they created reflected their understanding of environmental responsibility based on the Indic cultural values they had imbibed from their school and their communities.

## 3. Results and Discussion

At the end of each workshop class, we collected feedback from the participants. Of the 148 children who provided post workshop feedback on a 5 point Likert scale, 130 children said they loved the workshop and 14 children said they liked the workshop. 112 children said they learned a lot and 20 said they learned quite a bit. Almost all children (except one) said that they would tell others to come to this workshop if we conducted this again and everyone said that they would like to attend future workshops like this on robotics.

The story telling and animations allowed space for the children to be innovative and creative. A few children created their own personalized dioramas on environment conservation based on their personal beliefs. All the children were very enthusiastic when they presented their work to their parents. We will publish a full paper on our findings once we complete analyzing the data we collected from our pilot workshops.

## ACKNOWLEDGMENTS
The authors express their gratitude to Sri Mata Amritanandamayi Devi, world-renowned humanitarian and spiritual leader without whose guidance, undying support and ceaseless encouragement, this project would not have been possible. We would like to thank the graphics design team of AMMACHI labs for creating wonderful and easy to assemble stencils for creating the robotic puppets. Finally, we also thank the teachers and students at Amrita Vidyalayam schools for being recipients of this pilot.